\documentclass[smallextended,natbib,authoryear 
]{svjour3}       

\smartqed  

\usepackage{graphicx,color}

\usepackage{lineno}

\usepackage[colorlinks=true, breaklinks=true, pdfstartview=FitV, linkcolor=blue, citecolor=blue, urlcolor=red]{hyperref}

\journalname{Celestial Mechanics and Dynamical Astronomy}

\begin{document}

\title{Delaunay variables approach to the elimination of the perigee in Artificial Satellite Theory
}
\title{Delaunay variables approach to the elimination of the perigee}

\author{Martin Lara
         \and
        Juan F.~San-Juan
         \and \\
        Luis M.~L\'opez-Ochoa 
}


\institute{M.~Lara \at
              Columnas de Hercules 1, ES-11100 San Fernando, Spain, \\
              Tel.: +34-856-119449\\
              \email{mlara0@gmail.com}
           \and
           J.F.~San-Juan \at
             Departamento de Matem\'aticas y Computaci\'on, Universidad de La Rioja, Spain \\
              Tel.: +34-941-299440\\
              \email{juanfelix.sanjuan@unirioja.es}
           \and
           L.M.~L\'opez-Ochoa \at
             Departamento de Ingenier{\'\i}a Mec\'anica, Universidad de La Rioja, Spain \\
              Tel.: +34-941-299516\\
              \email{luis-maria.lopezo@unirioja.es}
}

\date{Received: date / Accepted: date 
}

\maketitle{}


\begin{abstract}
Analytical integration in Artificial Satellite Theory may benefit from different canonical simplification techniques, like the elimination of the parallax, the relegation of the nodes, or the elimination of the perigee. These techniques were originally devised in polar-nodal variables, an approach that requires expressing the geopotential as a Pfaffian function in certain invariants of the Kepler problem. However, it has been recently shown that such sophisticated mathematics are not needed if implementing both the relegation of the nodes and the parallax elimination directly in Delaunay variables. Proceeding analogously, it is shown here how the elimination of the perigee can be carried out also in Delaunay variables. In this way the construction of the simplification algorithm becomes elementary, on one hand, and the computation of the transformation series is achieved with considerable savings, on the other, reducing the total number of terms of the elimination of the perigee to about one third of the number of terms required in the classical approach.

\keywords{Canonical simplification \and Artificial Satellite Theory \and Delaunay variables \and elimination of the perigee \and Lie transformations}
\end{abstract}


\section{Introduction}

The traditional aim of Artificial Satellite Theory (AST) is to provide an analytical solution for the orbital motion of man-made objects about celestial bodies, and specifically around the Earth \citep{Brouwer1959,Kozai1959,DepritRom1970,Claes1980}. Ideally, this would require the computation of the general solution of a time-dependent, six-dimensional system of first-order differential equations. Despite there is no way to obtain such analytical solution in a general situation, it may be obtained for simple forces models, as is the case of Keplerian motion or the problem of two fixed centers \citep[see][for instance]{BrouwerClemence1961,Pars1965,Battin1999}. Hence, one must be satisfied with computing approximate analytical solutions that may apply to specific regions in phase space, within a restricted precision, and for a limited time interval ---the relaxation in the standards of the solution allowing for corresponding simplifications in the forces models to be dealt with. 
\par

The most important effect in AST is the central attraction of the Earth, a force that, when considered alone and focusing on bounded motion, confines the motion to Keplerian ellipses which are commonly defined by their orbital elements ---a successful set of integration constants containing plenty of physical insight. Other forces acting on artificial satellites, as, for instance, non-centralities of the geopotential, third-body attraction, or atmospheric drag, have the effect of slightly distorting the Keplerian orbit. Because of that, the Variation of Parameters (VOP) method, in which solutions to the motion of artificial satellites are given by the time-history of their osculating elements, is the natural approach to the problem \citep[see chap.~10 of][for instance]{Battin1999}.
\par

Simple forces models may, obviously, facilitate the search for analytical solutions to the VOP equations. Thus, for instance, lunisolar perturbations may be ignored in the propagation of the lower satellite orbits, while the effect of atmospheric drag is commonly neglected for the higher orbits. Besides, for Earth-like bodies, non-centralities associated to the second-order zonal harmonic cause the most important distortions in Keplerian motion, whereas all other gravitational harmonics produce higher-order effects (of the order of the square of the second order zonal harmonic). Because of this, the problem in which all harmonic coefficients of the geopotential are neglected except the second-order zonal harmonic coefficient ($J_2$), is called the \emph{main problem} of AST \citep{Brouwer1959}. However, one should be aware that, even under these simplifications, the dynamical system corresponding to the simplified forces model may remain non-integrable, as is precisely the case of the main problem \citep{IrigoyenSimo1993,CellettiNegrini1995}.
\par

Alternatively, the computation of approximate analytical solutions to the VOP equations can be approached by perturbation methods \citep[see][for instance]{SandersVerhulst1985,Nayfeh2004}. Furthermore, when the forces model deals only with conservative forces, the perturbation approach can take advantage of canonical transformation theory and Hamiltonian methods. In these cases, the Lie transformations approach \citep{Hori1966,Deprit1969}, which allows for the explicit computation of the formal solution, provides an efficient alternative to the classical Poincar\'e-von Zeipel method.\footnote{Modifications of the Poincar\'e method by \citet{vonzeipel1916} in order to deal with degenerate Hamiltonians, shaped the frame in which \citet{Brouwer1959} achieved his famous solution of AST. Hence, the method is sometimes known as the von Zeipel-Brouwer method \citep[cf.][]{Ferraz2007}}
\par

The common trend when computing Hamiltonian perturbations is to find a canonical transformation from osculating to \emph{mean} variables \citep[see chap.~14, para.~8 of][for instance]{BrouwerClemence1961}.\footnote{Approaches based on non-osculating elements are also possible \citep{Gurfil2004}.} This transformation converts the original Hamiltonian (in osculating variables) into a new one (in mean variables) which, up to a certain truncation order, is free from the higher frequencies of the motion. New canonical transformations may also allow remove the long-period effects so as to arrive at a Hamiltonian in action-angle variables which is cyclic in the angles \citep{Arnold1989}. Therefore, up to the truncation order of the perturbation theory, the actions are constant and the angles evolve at constant rates, with values that depend on the actions as derived from Hamilton equations. Then, the analytical solution is obtained by plugging the time solution for the action-angle variables into the transformation equations from action-angle to mean variables, which in turn are plugged into the transformation equations from mean to osculating variables. When the perturbation solution is approached by Lie transformations, all transformations are computed explicitly as also is the analytical solution, which, if written explicitly, would appear as a truncated power series in a determined small parameter ---the higher order of the expansion the more accurate solution is expected for time scales and initial conditions compatible with the lack of convergence of the method \citep[cf.][]{FerrerSanJuanAbad2007}. One should be aware, however, that this kind of analytical solution is only able to represent orbits with circulating perigee, therefore missing an important class of orbits in AST, as is the case of frozen orbits \citep[see][for instance]{CuttingFrautnickBorn1978,RosboroughOcampo1991,CoffeyDepritDeprit1994,Shapiro1996,LaraSanJuanLopezOchoa2013a}.
\par

Out of resonances, short-period effects are removed by averaging terms depending on the mean anomaly, a procedure that transforms the Hamiltonian into a normal form \citep[see Appendix 6 of][for instance]{Arnold1989} in which the Delaunay action has been converted into a formal integral of the averaged problem. Because the geopotential is an implicit function of the mean anomaly through the explicit appearance of the true anomaly, one soon finds trouble if the normalization is carried out in closed form because of the early appearance of the equation of the center in the first stages of the procedure \citep{Deprit1982}. However, the Lie transformations procedure is not limited exclusively to Hamiltonian normalization, and it can be used for just simplifying a given Hamiltonian in order to prepare it for a subsequent normalization \citep{DepritFerrer1989}, in this way helping in dealing with some of the equation of the center related issues.
\par

The paradigm of Hamiltonian simplification in AST is provided by the elimination of the parallax algorithm \citep{Deprit1981}, which, used as intermediate simplification in the normalization of the main problem, reduces dramatically the total number of terms in the solution series \citep{CoffeyDeprit1982}. The original formulation of the elimination of the parallax was presented in polar-nodal variables, also referred to as Whittaker or Hill variables, a canonical set that requires the use of certain invariants of the Kepler problem in the formulation of the procedure. In particular, the whole procedure relies on the properties of the \textit{semilatus rectum} and the projections of the eccentricity vector in the nodal frame. This fact seems to make mandatory a preliminary discussion of the feasibility of formulating the geopotential as a Pfaffian function, as well as the study of the concomitant properties of the Lie derivative when applied to this Pfaffian. However, it has been recently shown that such sophisticated mathematics are unnecessary, and the elimination of the parallax is straightforwardly achieved by the standard Lie transformations approach when using Delaunay variables \citep{LaraSanJuanLopezOchoa2014}.
\par

Analogously, the relegation algorithm \citep{DepritPalacianDeprit2001}, which is used in AST to pursue the elimination of tesseral terms in closed form and which was originally formulated in polar-nodal variables \citep{SegermanCoffey2000}, provides the greatest insight when formulated in Delaunay variables, notably showing the fundamental role played by the eccentricity in the sub-synchronous relegation algorithm, which makes its efficiency questionable when compared with the standard procedure based on usual expansions of elliptic motion \citep{LaraSanJuanLopezOchoa2013}.
\par

Another transformation which came in the wake of the elimination of the parallax and, probably because of this, was originally formulated in polar-nodal variables is the elimination of the perigee \citep{AlfriendCoffey1984}. This transformation is designed to remove trigonometric terms in the argument of the perigee which still remain in the Hamiltonian after the elimination of the parallax, and is based on the flexibility of the Lie transformations method in the computation of the generating function of the transformation, which is solved from the homological equation up to an additive arbitrary function pertaining to the kernel of the Lie derivative\footnote{Definitions related to the Lie transformations lingo may be consulted in the book of \citet{Ferraz2007}, for instance}. While the effect of using integration ``constants'' of this kind was not unknown at the times were the elimination of the perigee was devised \citep{Kozai1962,Morrison1965}, the clever way in which they are used in the elimination of the perigee definitely opens new ways in the construction of Lie transformations-based algorithms.
\par

The elimination of the perigee was proposed as an intermediate transformation that removes non-essential complications in a following normalization of the zonal problem, and has been successfully implemented in different software packages \citep{CoffeyAlfriend1984,SanJuan1994,SanJuan1998}. Besides, the usefulness of this canonical transformation is not restricted to the preparation of the zonal Hamiltonian for a subsequent Delaunay normalization. Indeed, the Hamiltonian resulting from the elimination of the perigee can be shaped in the form of a perturbed harmonic oscillator, which may be integrated by the Krylov-Bogoliubov-Mitropolsky method \citep{KrylovBogoliubov1947} thus providing an efficient alternative to the Delaunay normalization \citep{AbadSanJuanGavin2001}.
\par

One should note that the elimination of the perigee does not remove any polar-nodal variable from the Hamiltonian, but a Delaunay one. In consequence, when using polar-nodal variables the argument of the perigee must be decoupled from the argument of the latitude throughout the Lie transformations procedure, in which, besides, the dependence of the involved invariants in this angle must be carefully traced. However, it must be emphasized that the use of polar-nodal variables is more a matter of choice than a requirement of the simplification algorithm.
\par

This is  precisely the aim of the present research: To show how the reformulation of the elimination of the perigee algorithm using Delaunay variables releases this canonical simplification from unnecessary subtleties, thus making the Lie transformations procedure for eliminating the perigee elementary.
\par

In spite of the fact that the elimination of the perigee can be applied to a general zonal problem, the basic facts in the implementation of the algorithm are fully illustrated with the main problem. Therefore, this paper limits itself to this simpler model in the description of the algorithm. 
\par

The paper is organized as follows. First, the standard formulation of the main problem of AST in Delaunay variables is summarized, as well as the simplifications obtained after applying the elimination of the parallax transformation. The latter sets the main problem Hamiltonian in a convenient shape for approaching the elimination of the perigee. Next, a full description of the elimination of the perigee algorithm is provided. At difference from the original formulation in polar-nodal variables, this algorithm is reformulated in Delaunay variables. Far from being a trivial reformulation in a different set of canonical variables, the use of Delaunay variables clearly illustrates the value of adding arbitrary integration constants to the generating function of a perturbation theory by Lie transformations. Finally, a Delaunay normalization is carried out up to the fourth order of $J_2$, thus achieving the analytical perturbation solution without need of leaving the framework provided by the Delaunay variables in any of the three canonical transformations being carried out.
\par

\section{Preliminaries: The main problem} \label{s:potential}

When the main problem of AST is formulated in Cartesian coordinates and their conjugate momenta, the Hamiltonian depends on all the variables and, therefore, has three degrees of freedom. However, the simple formulation in spherical or cylindrical coordinates reveals the cylindrical symmetry of the problem which makes the polar component of the angular momentum $H$ an integral. The reduced problem has two degrees of freedom and is conservative; therefore, for each manifold $\mathcal{H}=$ constant its dynamics can be studied by means of the usual non-linear dynamics tools, such as Poincar\'e surfaces of section and periodic orbits \citep{Danby1968,Broucke1994}.
\par

\subsection{Hamiltonian in Delaunay variables}

Alternativelly, the fact that $J_2$ is small ---of the order of $10^{-3}$ in the case of the earth--- further allows us to set the main problem Hamiltonian in the terms of the following perturbation problem
\begin{equation} \label{initial}
\mathcal{H}=\mathcal{H}_{0,0}+\mathcal{H}_{1,0},
\end{equation}
where the first index of each Hamiltonian term notes the row and the second index the column in the construction of the ``Lie triangle'' \citep{Deprit1969}.
The term $\mathcal{H}_{0,0}$ is the Keplerian
\begin{equation}
\mathcal{H}_{0,0}=\frac{1}{2}(X^2+Y^2+Z^2)-\frac{\mu}{r},
\end{equation}
where $\mu$ is the Earth's gravitational parameter, $r=\sqrt{x^2+y^2+z^2}$ is the distance from the origin of coordinates, and $(x,y,z)$ and $(X,Y,Z)$ are position and velocity, respectively, in Cartesian coordinates.
The perturbation term $\mathcal{H}_{1,0}$ is
\begin{equation}
\mathcal{H}_{1,0}=J_2\,\frac{\mu}{r}\,\frac{\alpha^2}{r^2}\,P_2(\sin\varphi),
\end{equation}
where $\alpha$ is the mean equatorial radius of the Earth, $\varphi$ is the latitude, and $P_2$ is the Legendre polynomial of degree 2
\begin{equation}
P_2(\xi)\equiv-\frac{1}{2}+\frac{3}{2}\xi^2.
\end{equation}
\par

Orbital problems are conveniently described by the variation of their orbital elements $(a,e,i,\omega,\Omega,M)$, for semimajor axis, eccentricity, inclination, argument of the perigee, longitude of the ascending node, and mean anomaly, respectively. However, these variables are not canonical, so one must select a different set of variables in order to approach the problem of Hamiltonian perturbations. Furthermore, the selection of variables is essential in the computation of a perturbation theory, where the action-angle variables of the unperturbed problem are generally accepted as playing a fundamental role.
\par

The action-angle variables of the Kepler problem are materialized by the set of Delaunay canonical variables $(\ell,g,h,L,G,H)$, where $\ell=M$, $g=\omega$, $h=\Omega$, the Delaunay action $L=\sqrt{\mu\,a}$ is the conjugate momentum to $\ell$, the modulus of the angular momentum $G=L\,\sqrt{1-e^2}$ is the conjugate momentum to $g$, and the polar component of the angular momentum $H=G\cos{i}$ is the conjugate momentum to $h$. However, it is worth noting that the main problem Hamiltonian cannot be explicitly written in terms of Delaunay variables because the conic solution \citep[see Eq.~44, chap.~1 of][for instance]{BrouwerClemence1961}
\begin{equation} \label{conic}
r=\frac{a\,(1-e^2)}{1+e\cos{f}},
\end{equation}
involves the true anomaly $f$ which is an implicit function of the mean anomaly $f=f(M;e)=f(\ell;G,L)$ through the Kepler equation
\begin{equation}
\ell=u-e\sin{u},
\end{equation}
where the eccentric anomaly $u$ is related to the true one by the relation
\begin{equation}
\tan\frac{u}{2}=\sqrt{\frac{1-e}{1+e}}\tan\frac{f}{2}.
\end{equation}
\par

In spite of different expansions of the elliptic motion appearing in the literature \citep[see chap.~2 of][for instance]{BrouwerClemence1961}, these kinds of expansions must be obviously avoided in the computation of a closed-form theory. Because of this, as well as for the direct physical insight of orbital elements, the main problem Hamiltonian is rewritten as
\begin{eqnarray} \label{H0}
\mathcal{H}_{0,0} &=& -\frac{\mu}{2a}, \\ \label{H1}
\mathcal{H}_{1,0} &=&
-J_2\,\frac{\mu}{2r}\,\frac{\alpha^2}{r^2}\left[1-\frac{3}{2}\sin^2i+\frac{3}{2}\sin^2i\cos(2f+2\omega)\right],
\end{eqnarray}
where all symbols: $a$, $r$, $i$, $f$ and $\omega$, must be understood as \emph{functions} of Delaunay variables. Besides, it is convenient for what follows to introduce the eccentricity function
\begin{equation}
\eta=\sqrt{1-e^2}=G/L,
\end{equation}
the Keplerian mean motion
\begin{equation}
n=\sqrt{\mu/a^3}=\mu^2/L^3,
\end{equation}
and the usual abbreviations
\begin{equation}
c\equiv\cos{i}=H/G, \qquad s\equiv\sin{i}=\sqrt{1-H^2/G^2}.
\end{equation}

\subsection{Perturbations by Lie transformations}

Given a Hamiltonian of the form $\mathcal{H}=\mathcal{H}_{0,0}+\varepsilon\,\mathcal{D}$, where $\mathcal{H}_{0,0}$ is integrable, $\varepsilon$ is a small parameter, and the disturbing function $\mathcal{D}=(1/\varepsilon)\sum_{m\ge1}(\varepsilon^m/m!)\,\mathcal{H}_{m,0}$ is given by its expansion in power series of $\varepsilon$, Deprit's perturbation theory by Lie transformations \citep{Deprit1969,Ferraz2007} enables the computation of a canonical transformation from old (initial) to new (transformed) variables such that the Hamiltonian is  simplified when expressed in the new variables $\mathcal{H}'=\sum_{m\ge0}(\varepsilon^m/m!)\,\mathcal{H}_{0,m}$. The canonical transformation is derived from a generating function $W$ which is also given by its expansion in powers of the small parameter, namely $W=\sum_{m\ge0}(\varepsilon^m/m!)\,W_{m+1}$.
\par

The Lie transformations method is based on the recurrence
\begin{equation} \label{triangle}
\mathcal{H}_{n,q}=\mathcal{H}_{n+1,q-1}+
\sum_{0\le{m}\le{n}}{n\choose{m}}\,\{\mathcal{H}_{n-m,q-1};W_{m+1}\},
\end{equation}
commonly dubbed as Deprit's triangle, where $\{F_1;F_2\}$ notes the Poisson bracket of the functions $F_1$ and $F_2$ expressed in some set of canonical variables. At each step $m$ of the procedure, Eq.~(\ref{triangle}) gives rise to the so-called homological equation
\begin{equation} \label{homom}
\mathcal{L}_0(W_m)=\widetilde\mathcal{H}_{m,0}-\mathcal{H}_{0,m},
\end{equation}
where, in a perturbation approach, the new Hamiltonian term $\mathcal{H}_{0,m}$ and the corresponding term of the generating function $W_m$ are unknown, and the Lie operator $\mathcal{L}_0$ is a Poisson bracket operator: $\mathcal{L}_0(F_1)=\{F_1,\mathcal{H}_{0,0}\}$, where $F_1$ is some function of the canonical variables. Terms $\widetilde\mathcal{H}_{m,0}$ in Eq.~(\ref{homom}) are known from previous computations derived from successive evaluations of Eq.~(\ref{triangle}), whereas the term $\mathcal{H}_{0,m}$ is chosen at our convenience. Finally, the solution of the partial differential equation $(\ref{homom})$ gives $W_m$.
\par

In particular, when dealing with perturbed Keplerian problems in Delaunay variables, $\mathcal{H}_{0,0}$ depends only on the Delaunay action $L$, cf.~Eq.~(\ref{H0}). Therefore, the Lie operator is simply
\begin{equation}
\mathcal{L}_0\equiv{n}\,\frac{\partial\;}{\partial\ell}.
\end{equation}
Besides, the notation
\begin{equation} \label{original2prime}
(\ell,g,h,L,G,H)\stackrel{\mathcal{T}_1}\longrightarrow(\ell',g',h',L',G',H').
\end{equation}
is used for representing the transformation from old to new (prime) variables; in particular $\mathcal{T}_1:F$ notes the result of applying the transformation to the function $F$ in order to express it in the prime variables.
\par

In the process of normalizing the main problem Hamiltonian in closed form, it happens that short-period terms related to the appearance of $1/r^3$ in Eq.~(\ref{H1}) introduce the equation of the center (the difference between the true and the mean anomaly) in the first stages of the Lie transformations procedure, a fact that makes it difficult to progress in the perturbation theory to higher orders. However, the normalization can be preceded by a preparatory simplification which eliminates these kinds of terms, thus notably easing the normalization process.
\par

Specifically, the elimination of the parallax \citep{Deprit1981} is a Lie transformation which has the effect of removing all short-period effects from the Hamiltonian (in the new variables and up to the truncation order) except for the coefficient $1/r^2$. This simplification shows notable merits, reducing the number of terms needed in the implementation of a third-order solution to the main problem to one fourth of those required by a direct normalization of the problem \citep{CoffeyDeprit1982}. For the sake of completeness, the recent approach to the elimination of the parallax proposed in \citep{LaraSanJuanLopezOchoa2014} is summarized in what follows.

\subsection{The main problem Hamiltonian after the elimination of the parallax}

Reduction of parallactic terms $1/r^m$ with $m>2$, is based on the identity
\begin{equation}\label{identity}
r^m=r^2\,r^{m-2}=r^2\left(\frac{a\,\eta^2}{1+e\cos{f}}\right)^{\!m-2},\qquad m>2.
\end{equation}
which is immediately derived from  Eq.~(\ref{conic}). Then, using Eq.~(\ref{identity}), Eq.~(\ref{H1}) is rewritten in the form
\begin{eqnarray} \label{mainH1p}
\mathcal{H}_{1,0} &=& 
-\frac{\mu}{2a\,\eta^2}\,\frac{\alpha^2}{r^2}\,\frac{1}{2}J_2\Big[(2-3s^2)\,(1+e\cos{f}) \\ \nonumber
&& +\frac{3}{2}s^2\,e\cos(f+2\omega)+3s^2\cos(2f+2\omega)+\frac{3}{2}s^2\,e\cos(3f+2\omega)
\Big].
\end{eqnarray}

From Eq.~(\ref{homom}), the first order of Deprit's perturbation theory is
\begin{equation}\label{homom1st}
\mathcal{L}_0(W_1)=\{W_1,\mathcal{H}_{0,0}\}=
n\,\frac{\partial{W}_1}{\partial\ell}=\widetilde\mathcal{H}_{1,0}-\mathcal{H}_{0,1}.
\end{equation}
where $\widetilde\mathcal{H}_{1,0}=\mathcal{H}_{1,0}$ from the first order of Deprit's triangle in Eq.~(\ref{triangle}). Then, at difference from the usual averaging over the \emph{mean} anomaly $\ell$, $\mathcal{H}_{0,1}$ is chosen by removing terms in Eq.~(\ref{mainH1p}) in which the \emph{true} anomaly $f$ appears explicitly. Thus,
\begin{equation}\label{h01}
\mathcal{H}_{0,1}=-\frac{\mu}{2a}\,J_2\,\frac{1}{\eta^2}\,\frac{\alpha^2}{r^2}\left(1-\frac{3}{2}s^2\right).
\end{equation}
Remark that, albeit the explicit appearance of the true anomaly has been completely removed from Eq.~(\ref{mainH1p}), $\mathcal{H}_{0,1}\ne\langle\mathcal{H}_{1,0}\rangle_f$, where $\langle\cdot\rangle_f$ means averaging over the true anomaly, because it still depends on $r$, and $r\equiv{r}(f)$ as is evident in Eq.~(\ref{conic}). Therefore, the new Hamiltonian term in Eq.~(\ref{h01}) is affected of long-period terms as well as remaining short-period terms.
\par

Note that the new Hamiltonian term $\mathcal{H}_{0,1}$ should be written in new (prime) variables because it is the result of the canonical transformation in Eq.~(\ref{original2prime}). However,  for simplicity we avoid the prime notation throughout this paper as far as there is not risk of confusion.
\par

Then, $W_1$ is solved from Eq.~(\ref{homom1st}) by a simple quadrature, which is carried out in closed form of the eccentricity based on the standard relation between the true and mean anomalies
\begin{equation} \label{dldf}
a^2\,\eta\,\mathrm{d}\ell=r^2\,\mathrm{d}f,
\end{equation}
which is derived from the preservation of the angular momentum in the Kepler problem \citep[see chap.~1, para.~7 of][for instance]{BrouwerClemence1961}. Namely, combining Eqs.~(\ref{homom1st}) and (\ref{dldf}),
\begin{equation}
W_1=\frac{1}{n}\int(\mathcal{H}_{1,0}-\mathcal{H}_{0,1})\,\mathrm{d}\ell
=\frac{1}{n}\int(\mathcal{H}_{1,0}-\mathcal{H}_{0,1})\,\frac{r^2}{a^2\,\eta}\,\mathrm{d}f,
\end{equation}
which results in
\begin{eqnarray}\label{W1}
W_1 &=&-n\,\alpha^2\,\frac{J_2}{8 \eta ^3}\Big[(4-6s^2)\,e\sin{f} \\ \nonumber
&& +3s^2\,e\sin(f+2\omega)+3s^2\sin(2f+2\omega)+s^2\,e\sin(3f+2\omega)\Big].
\end{eqnarray}
\par

A new evaluation of Eq.~(\ref{homom}) provides the homological equation of the second order
\begin{equation}\label{homo2}
\mathcal{L}_0(W_2)=\widetilde\mathcal{H}_{2,0}-\mathcal{H}_{0,2},
\end{equation}
where
\begin{equation}
\widetilde\mathcal{H}_{2,0}=\{\mathcal{H}_{0,1},W_1\}+\{\mathcal{H}_{1,0},W_1\}.
\end{equation}
After computing the Poisson brackets and using Eq.~(\ref{identity}) for eliminating parallactic terms $(1/r)^m$ with $m>2$, we obtain
\begin{eqnarray}
\widetilde\mathcal{H}_{2,0} &=& -\frac{\mu}{2a}\,\frac{\alpha^2}{a^2}\,\frac{\alpha^2}{r^2}\,J_2^2\,\frac{1}{\eta^6}\left\{
\frac{5}{2}
+\frac{3}{4}e^2
+\left(4-\frac{27}{4}s^2+\frac{27}{16}s^4\right)e\cos{f} \right.
\\ \nonumber
&& +\left(\frac{3}{4}-\frac{3}{4}s^2-\frac{15}{32}s^4\right)e^2\cos2f
  -s^2\left[\frac{21}{4}-\frac{3}{4}e^2  \right.
\\ \nonumber
&&  + \left(\frac{21}{8}-\frac{45}{16}s^2\right)e^2\cos2\omega
    +\left(\frac{77}{8}-\frac{21}{2}s^2\right)e\cos(f+2\omega)
\\ \nonumber
&&  +\left(5-\frac{21}{4}s^2+\frac{3}{8}\left(2-s^2\right)e^2\right)\cos(2f+2\omega)
\\ \nonumber
&&  \left.
    -\left(\frac{15}{8}-3s^2\right)e\cos(3f+2\omega)
    -\left(\frac{15}{8}-\frac{39}{16}s^2\right)e^2\cos(4f+2\omega)
    \right]
\\ \nonumber
&& +s^4\left[
   \frac{21}{8}-\frac{15}{32}e^2+\frac{15}{64}e^2\cos(2f+4\omega) +\frac{9}{32}e\cos(3f+4\omega)
\right.
\\ \nonumber
&& \left.\left. 
   -\left(\frac{3}{8}-\frac{3}{32}e^2\right)\cos(4f+4\omega)
   -\frac{15}{32}e\cos(5f+4\omega)
   -\frac{9}{64}e^2\cos(6f+4\omega)
\right]
\right\},
\end{eqnarray}
and choose $\mathcal{H}_{0,2}$ like before: by removing the explicit appearance of $f$ in $\widetilde\mathcal{H}_{2,0}$. We get
\begin{eqnarray}
\mathcal{H}_{0,2} &=& -\frac{\mu }{2 a}\,\frac{\alpha^2}{a^2}\,\frac{\alpha^2}{r^2}\,J_2^2\,\frac{1}{8 \eta ^6}\left[
\frac{5}{2}-\frac{21}{4}s^2 +\frac{21}{8}s^4 \right. \\ \nonumber
&& \left. +\left(\frac{3}{4}-\frac{3}{4}s^2-\frac{15}{32}s^4\right)e^2
-s^2\left(\frac{21}{8}-\frac{45}{16}s^2\right)e^2\cos2\omega\right].
\end{eqnarray}
As in the first order case, the homological equation, Eq.~(\ref{homo2}), is solved by quadrature to give the second order term of the generating function $W_2$.
\par

The elimination of the parallax in Delaunay variables can be extended to any order of the perturbation method without leaving the algebra of trigonometric functions. Thus, after eliminating the parallax up to the fourth order of $J_2$, the new Hamiltonian $\mathcal{P}=\mathcal{T}_1:\mathcal{H}$ is given by the Poisson series
\begin{equation} \label{parallax}
\mathcal{P}=-\frac{\mu}{2a}
-\frac{\mu}{2a}\,\frac{1}{\eta^2}\,\frac{\alpha^2}{r^2}\,
\sum_{i=1}^4\frac{J_2^i}{i!}\left(\!\frac{\alpha}{a\,\eta^2}\!\right)^{\!\!2i-2}\,
\sum_{j=0}^m\left[\sum _{k=0}^{m-j}e^{2 k} q_{i,j,k}(s)\right]e^{2j}s^{2j}\cos(2j\,\omega),
\end{equation}
where $m=i/2$ is an integer division, and the inclination polynomials $q_{i,j,k}$ are presented in Table \ref{t:qijk} of the Appendix \citep[cf.][]{CoffeyDeprit1982,LaraSanJuanLopezOchoa2014}. Note that the symbols $a$, $e$, $\eta$, $r$, $s$, and $\omega$ are now functions of prime variables, since they depend on the Delaunay transformed variables.
\par

In spite of the fact that $\mathcal{P}$ has the same degrees of freedom as the original Hamiltonian $\mathcal{H}$, and may seem much more complex, the new Hamiltonian has been simplified in the sense that $\mathcal{P}$ only depends on short-period terms through the factor $1/r^2$, whereas $\mathcal{H}$ involves $1/r^3$.
\par

The Lie transformations method also provides the generating function of the transformation $\mathcal{T}_1$, from which the explicit transformation equations
\begin{equation}
\xi=\xi(\ell',g',h',L',G',H'),\qquad
\xi\in(\ell,g,h,L,G,H),
\end{equation}
are obtained by means of the straightforward application of the method to the Delaunay variables themselves. Full details on the elimination of the parallax algorithm can be consulted in the original reference of \citet{Deprit1981} where the canonical simplification is formulated in polar-nodal variables, or in our own reformulation of the procedure in Delaunay variables proposed in \citep{LaraSanJuanLopezOchoa2014}. The latter, in addition to the simplicity of the derivations, reduces the number of terms to evaluate in the transformation to about the 70\% of the number of terms required by the former, as illustrated in Table \ref{t:paralax}.

\begin{table*}[htbp]
\caption{Number of terms of the (expanded) Hamiltonian and generating function of the elimination of the parallax in Delaunay ($H$, $W$) and polar-nodal variables ($K$, $V$).}
\label{t:paralax}
\centering 
\begin{tabular}{@{}crrcrrc@{}}
\multicolumn{1}{c}{$m$} & \multicolumn{1}{c}{$H_m$} & \multicolumn{1}{c}{$K_m$} & \multicolumn{1}{c}{$1-H_m/K_m$} & \multicolumn{1}{c}{$W_m$} & \multicolumn{1}{c}{$V_m$} & \multicolumn{1}{c}{$1-W_m/V_m$} \\
\hline\noalign{\smallskip}
    1 &  2 &  2 & 0 &   5 &   7 & 28.57\% \\
    2 &  8 &  8 & 0 &  22 &  32 & 31.25\% \\
    3 & 11 & 12 & 8.33\% &  73 & 112 & 34.82\% \\
    4 & 30 & 30 & 0 & 180 & 264 & 31.82\% \\
total & 51 & 52 & 1.92\% & 280 & 415 & 32.53\% \\
\noalign{\smallskip}\hline
\end{tabular}
\end{table*}

\section{The elimination of the perigee} \label{s:disturbing}

Prior to normalizing the remaining short period terms in Eq.~(\ref{parallax}), \citet{AlfriendCoffey1984} propose first removing long-period terms depending on the argument of the perigee from the Hamiltonian. This is done by a new Lie transformation
\begin{equation} \label{prime2second}
(\ell',g',h',L',G',H')\stackrel{\mathcal{T}_2}\longrightarrow(\ell'',g'',h'',L'',G'',H''),
\end{equation}
which, up to the truncation order, converts the modulus of the angular momentum vector in a formal integral of the perturbation problem in the double prime variables. 
\par

At the first order, the term $\widetilde\mathcal{H}_{1,0}$ is given by the first order of $J_2$ in Eq.~(\ref{parallax}), namely
\begin{equation}
\widetilde\mathcal{H}_{1,0}=-\frac{\mu}{2a}\frac{\alpha^2}{r^2}\,\frac{J_2}{\eta^2}\left(1-\frac{3}{2}s^2\right).
\end{equation}
Because $\widetilde\mathcal{H}_{1,0}$ does not depend on the argument of the perigee ($\omega\equiv{g}''$), $\mathcal{H}_{0,1}$ is chosen as $\mathcal{H}_{0,1}=\widetilde\mathcal{H}_{1,0}$. In consequence, at the first order, Eq.~(\ref{homom}) results in $\mathcal{L}_0(W_1)=0$, which is trivially solved by quadrature to give
\begin{equation}
W_1=V_1(-,g'',-,L'',G'',H''),
\end{equation}
where a dash emphasizes the absence of the corresponding variable. The ``constant'' $V_1$ should be periodic in $g''$ to prevent the introduction of secular terms in the generating function, but otherwise can be taken as arbitrary at this step. However, the adequate selection of $V_1$ is crucial to the elimination of the perigee in second- and higher-order steps of this algorithm, in a parallel way as \citet{AlfriendCoffey1984} did in the case of polar-nodal variables.
\par

\subsection{Second order}

The homological equation of the second order is again Eq.~(\ref{homo2}) where, now
\begin{equation} \label{tildeH02}
\widetilde\mathcal{H}_{2,0}=\{\mathcal{H}_{0,1},W_1\}+\{\mathcal{H}_{1,0},W_1\}+\mathcal{H}_{2,0}.
\end{equation}
That is:
\begin{eqnarray} \label{H02t}
\widetilde\mathcal{H}_{2,0} &=&
-\frac{\mu}{2a}\,\frac{\alpha^2}{r^2}\frac{J_2}{\eta^5}\,(1-3 c^2)\,(2e\sin{f}+e^2\sin2f)\,\frac{\partial{V}_1}{\partial{L}''}
\\ \nonumber 
&& +\frac{\alpha^2}{r^2}\frac{J_2}{\eta^3}
\left[ \frac{3}{2}\,(1-5c^2)+(1-3c^2)\left(\frac{1}{e}\cos{f}+\frac{1}{2}\cos2f\right) \right]n\,\frac{\partial{V}_1}{\partial{g}''} \qquad
\\ \nonumber
&& +\frac{\mu}{2a}\frac{\alpha^2}{r^2}\,\frac{\alpha^2}{a^2}\,\frac{J_2^2}{\eta^6} \left[
\frac{1}{8}(1-21c^4) 
+\frac{3}{32}(5-18 c^2+5c^4)\,e^2 
\right. \\ && \left. \nonumber
-\frac{3}{16}(1-15c^2)\,e^2 s^2\cos2\omega
\right].
\end{eqnarray}
Note the mixed notation in Eq.~(\ref{H02t}): while the partial derivatives are expressed in function of the (double prime) Delaunay variables, the orbital elements notation is used for the rest of the terms of $\widetilde\mathcal{H}_{2,0}$; in particular, we recall that, now, $\omega\equiv\omega(g'')=g''$.
Besides, $e\ne0$ is implicitly assumed in Eq.~(\ref{H02t}) for the known singularities of orbital elements and their canonical counterpart provided by Delaunay variables. However, this does not mean to restrict the applicability of the theory to the case of non-circular orbits. Indeed, in the fashion of \citet{Lyddane1963}, after the theory has been computed, it can be reformulated in non-singular elements ---without need of restricting to the canonical case \citep[cf.][]{DepritRom1970}.
\par

It worths noting that, in the same way as it is done in the parallax elimination, after computing Eq.~(\ref{tildeH02}) the identity in Eq.~(\ref{identity}) has been used to avoid the explicit appearance of inverse powers of $r$ higher than quadratic in Eq.~(\ref{H02t}).
\par

Since the aim of this Lie transformation is to remove the perigee from the new Hamiltonian, the new Hamiltonian term $\mathcal{H}_{0,2}$ is chosen to comprise only those terms of $\widetilde\mathcal{H}_{2,0}$ that are free from $\omega\equiv{g}''$. Then, taking into account that the partial derivatives of $V_1$ in Eq.~(\ref{H02t}) are periodic in $g''$ because of the corresponding periodicity of $V_1$, 
\begin{equation}
\mathcal{H}_{0,2}=
\frac{\mu}{2a}\frac{\alpha^2}{r^2}\,\frac{\alpha^2}{a^2}\,\frac{J_2^2}{\eta^6}
\left[ \frac{1}{8}(1-21c^4) +\frac{3}{32}(5-18 c^2+5c^4)\,e^2 \right],
\end{equation}
which has been obtained by neglecting periodic functions of $\omega$ in Eq.~(\ref{H02t}).

Therefore, Eq.~(\ref{homo2}) is written
\begin{eqnarray} \label{dW2l}
n\,\frac{\partial{W}_2}{\partial\ell''} &=&
-\frac{\mu}{2a}\,\frac{\alpha^2}{r^2}\frac{J_2}{\eta^5}\,(1-3 c^2)\,(2e\sin{f}+e^2\sin2f)\,\frac{\partial{V}_1}{\partial{L}''}
\\ \nonumber &&
+\frac{\alpha^2}{r^2}\frac{J_2}{\eta^3}
\left[ \frac{3}{2}\,(1-5c^2)+(1-3c^2)\left(\frac{1}{e}\cos{f}+\frac{1}{2}\cos2f\right) \right]n\,\frac{\partial{V}_1}{\partial{g}''}
\\ \nonumber
&& -\frac{\mu}{2a}\frac{\alpha^2}{r^2}\,\frac{\alpha^2}{a^2}\,\frac{J_2^2}{\eta^6}\,\frac{3}{16}(1-15c^2)\,e^2 s^2\cos2\omega,
\end{eqnarray}
which can be integrated in closed form of the eccentricity by a simple quadrature. Indeed, based on Eq.~(\ref{dldf}), Eq.~(\ref{dW2l}) is rewritten\footnote{Note that, after the different transformations, the transformed Hamiltonian continues to represent perturbed Keplerian motion, and hence the usual relations of elliptic motion still apply in spite of the different meaning of the prime variables from the original ones.}
\begin{eqnarray}
\frac{\partial{W}_2}{\partial{f}} &=&
 -n\,\alpha^2\,\frac{\alpha^2}{a^2}\,\frac{J_2^2}{\eta^7}\,\frac{3}{32}(1-15c^2)\,e^2 s^2\cos2\omega \\ \nonumber
&& -n\,\alpha^2\,\frac{J_2}{\eta^6}\,(1-3 c^2)\left(e\sin{f}+\frac{1}{2}\,e^2\sin2f\right)\frac{\partial{V}_1}{\partial{L}''} \\ \nonumber
&&+\frac{\alpha^2}{a^2}\frac{J_2}{\eta^4}
\left[ \frac{3}{2}\,(1-5c^2)+(1-3c^2)\left(\frac{1}{e}\cos{f}+\frac{1}{2}\cos2f\right) \right]\frac{\partial{V}_1}{\partial{g}''}
\end{eqnarray}
and, hence,
\begin{eqnarray} \label{W2}
W_2 &=&
f\,\frac{3}{2}\,\frac{\alpha^2}{a^2}\,\frac{J_2}{\eta^4}\left[(1-5c^2)\frac{\partial{V}_1}{\partial{g}''}
-\frac{n\,\alpha^2}{2}\,\frac{J_2}{\eta^3}\,\frac{1}{8}(1-15c^2)\,e^2 s^2\cos2\omega
\right]
\\ \nonumber
&& +\frac{n\,\alpha^2}{2}\,\frac{J_2}{\eta^6}\,(1-3 c^2)\left(2e\cos{f}+\frac{1}{2}e^2\cos2f\right)\frac{\partial{V}_1}{\partial{L}''}
\\ \nonumber
&& +\frac{\alpha^2}{a^2}\frac{J_2}{\eta^4}
(1-3c^2)\left(\frac{1}{e}\sin{f}+\frac{1}{4}\sin2f\right)\frac{\partial{V}_1}{\partial{g}''} +V_2(-,g'',-,L'',G'',H'').
\end{eqnarray}
\par

Now, in order to prevent the appearance of secular terms in Eq.~(\ref{W2}), $V_1$ is chosen in such a way that it makes null the coefficient of $f$ in Eq.~(\ref{W2}). This is easily achieved by solving a simple quadrature, which provides
\begin{equation} \label{V1}
V_1=\frac{1}{32}\,n\,\alpha^2\,\frac{J_2}{\eta^3}\,\frac{1-15c^2}{1-5c^2}\,e^2 s^2\sin2\omega.
\end{equation}
\par

Finally, by replacing Eq.~(\ref{V1}) into Eq.~(\ref{W2}), one gets
\begin{equation} \label{W2a}
W_2 = \frac{J_2^2\,n\,\alpha^4}{16a^2\eta^7}\,s^2\,(1-15c^2)
\frac{1-3c^2}{1-5c^2}\left[e\sin(f+2\omega)+\frac{e^2}{4}\sin(2f+2\omega)\right]+V_2,
\end{equation}
where the integration ``constant'' $V_2$ needs to wait until the following step of the Lie transformations procedure in order to be determined.
\par

It deserves to mention that $V_1$ is determined except for a new arbitrary ``constant'', which must be independent of the angle $\omega\equiv{g}''$ but may depend on the momenta. Introduction of a constant of this type would affect the computation of $\partial{V}_1/\partial{L}''$ in Eq.~(\ref{W2}), in this way making the generating function more intricate by adding to $W_2$ new terms depending on $\cos{f}$ and $\cos2f$. The possible benefits of using such a constant have not been explored in view of the simplicity of Eq.~(\ref{W2a}). Indeed, in addition to preventing the appearance of secular terms in $W_2$, Eq.~(\ref{V1}) has the virtue of canceling the combinations in $\sin(f-2\omega)$ and $\sin(2f-2\omega)$ resulting from the evaluation of $\partial{V}_1\partial{L}''$ and $\partial{V}_1\partial{g}''$, thus entailing a reduction in the number of terms of the generating function in this step, which also affects the next ones.
\par

Note the coefficient $1-5c^2$ appearing in Eq.~(\ref{W2a}) as a divisor, thus preventing the use of the theory close to the inclination given by $\cos^2i=1/5$. This deficiency in the elimination of the perigee transformation was expected from the essential singular character of the critical inclination \citep[see][and references therein]{Cushman1983,CoffeyDepritMiller1986}.
\par

Remind that the symbols $n$, $a$, $e$, $\eta$, $r$, $c$, $s$, and $\omega$ in preceeding equations are now functions of the double prime Delaunay variables.
\par

\subsection{Higher orders}

The procedure can be continued up to any desired order. Indeed, proceeding as before, one finds the Hamiltonian $\mathcal{Q}=\mathcal{T}_2:\mathcal{P}$ given by
\begin{equation} \label{perigee}
\mathcal{Q}=-\frac{\mu}{2a}
-\frac{\mu}{2a}\,\eta^2\,\frac{a^2}{r^2}\,
\sum_{i=1}^4\frac{J_2^i}{i!}\left(\!\frac{\alpha}{a\,\eta^2}\!\right)^{\!\!2i}\,
\sum_{j=0}^{i-1}\left(\frac{e^2}{4-5s^2}\right)^{j} q_{i,j}(s),
\end{equation}
where $a$, $e$, $\eta$, $r$, and $s$ are expressed in double prime Delaunay variables, and the inclination polynomials $q_{i,j}$ are given in Table \ref{t:qij} of the Appendix up to the fourth order of $J_2$. Note that polynomials $q_{i,0}$ are the same as corresponding ones $q_{i,0,0}$ of Table \ref{t:qijk}, and therefore are not presented in Table \ref{t:qij}. Besides, for the convenience of interested readers in checking their own depelopments, specific expressions of $W_3$, and the integration ``constants'' $V_2$ and $V_3$ are also provided in the Appendix.
\par

Finally, the computational effort required by the elimination of the perigee in Delaunay and polar-nodal variables is illustrated in Table \ref{t:perigee}, which shows that the computation of higher orders of the elimination of the perigee simplification is much more efficient when using the Delaunay variables approach. Indeed, the number of terms of the transformation series of the elimination of the perigee in Delaunay variables is less than one third of the number of terms required by the original, polar-nodal approach. Obviously, if the transformation series in polar-nodal variables are rewritten in Delaunay variables, or vice-versa, both theories match. By verifying this fact, we obtained an important check on the correction of our results.

\begin{table*}[htbp]
\caption{Number of terms of the (expanded) Hamiltonian and generating function of the elimination of the perigee in Delaunay ($H$, $W$) and polar-nodal variables ($K$, $V$).}
\label{t:perigee}
\centering 
\begin{tabular}{@{}crrcrrc@{}}
\multicolumn{1}{c}{$m$} & \multicolumn{1}{c}{$H_m$} & \multicolumn{1}{c}{$K_m$} & \multicolumn{1}{c}{$1-H_m/K_m$} & \multicolumn{1}{c}{$W_m$} & \multicolumn{1}{c}{$V_m$} & \multicolumn{1}{c}{$1-W_m/V_m$} \\
\hline\noalign{\smallskip}
    1 &  2 &   2 & 0 &   2 &    2 &  \\
    2 &  6 &  15 & 60.00\% &  20 &   26 & 23.08\% \\
    3 & 14 &  83 & 83.13\% & 126 &  306 & 58.82\% \\
    4 & 25 & 318 & 92.14\% & 491 & 1659 & 70.40\% \\
Total & 47 & 418 & 88.76\% & 639 & 1993 & 67.94\% \\
\noalign{\smallskip}\hline
\end{tabular}
\end{table*}

\section{Analytic perturbation solution} \label{s:experiments}

Finally, the normalization of the Hamiltonian is achieved by the standard Delaunay normalization \citep{Deprit1982}. That is, a new canonical transformation
\begin{equation} \label{second2third}
(\ell'',g'',h'',L'',G'',H'')\stackrel{\mathcal{T}_3}\longrightarrow(\ell''',g''',h''',L''',G''',H'''),
\end{equation}
which removes the remaining short-periodic effects from Eq.~(\ref{perigee}), leading to $\mathcal{N}=\mathcal{T}_3:\mathcal{Q}$ given by
\begin{equation} \label{normalized}
\mathcal{N}=-\frac{\mu}{2a}
-\frac{\mu}{2a}\,
\sum_{i=1}^4\frac{J_2^i}{i!}\left(\!\frac{\alpha}{a\,\eta^2}\!\right)^{\!\!2i}\,
\frac{1}{(1-5c^2)^{i-1}}\,\sum_{j=0}^{2i-2}\,\eta^{j+1}\,p_{i,j}(s),
\end{equation}
where $a$, $e$, $\eta$ and $s$ are functions of the triple prime variables which only depend on momenta. The final normalization has been computed in closed form up to the fourth order of $J_2$ with ``MathATESAT'' \citep{SanJuanLopezOchoaLopez2011}. Corresponding inclination polynomials $p_{i,j}$ are given in Table \ref{t:qi}.
\par

Note that the appearance of the usual function $\beta=1/(1+\eta)$, which is sometimes used in normalization procedures that do not stem from the elimination of the perigee \citep{PalacianSanJuanYanguas1997,Healy2000}, is completely avoided throughout the whole procedure (simplification and normalization). Because of that, the normalized Hamiltonian can be written in the simple form of Eq.~(\ref{normalized}), which, if expanded, amounts to only 55 rational coefficients of the order of $J_2^4$, thus providing a notable economy when compared with the 144 coefficients of the order of $J_2^4$ needed when the secular Hamiltonian is organized in the form given by \citet{CoffeyDeprit1982}. 
\par

Then, after truncation, Eq.~(\ref{normalized}) is fully normalized: the actions $L'''$, $G'''$, and $H'''$ are integrals of the normalized motion, whereas the angles $\ell'''$, $g'''$, and $h'''$ evolve with constant rates which are derived from Hamilton equations
\begin{equation} \label{frequencies}
n_\ell=\frac{\partial\mathcal{N}}{\partial{L}'''}, \qquad
n_g=\frac{\partial\mathcal{N}}{\partial{G}'''}, \qquad
n_h=\frac{\partial\mathcal{N}}{\partial{H}'''}.
\end{equation}
\par

Lastly, the analytical solution is obtained by plugging the trivial integration of Eq.~(\ref{frequencies}) into the transformation equations from triple to double prime variables, which in turn are plugged into the transformation equations from double prime to prime variables, which in turn are substituted into the transformation equations from prime to original variables.

\section{Conclusions}

Artificial Satellite Theory is a major mathematical problem in aerospace engineering which requires the apparatus of perturbation theory to achieve useful analytical solutions. When dealing with geopotential perturbations, the Lie transformations method provides an excellent framework for progressing in the computation of higher orders of the solution, a procedure in which the use of canonical simplifications comes out as an essential step in the computation of a fully normalized, zonal Hamiltonian in closed form. However, these specialized applications of the Lie transformations method were originally formulated in polar-nodal variables, a fact that introduces non-essential subtleties in the respective algorithms. 
\par

The outcome of the present research supports previous results of the authors which demonstrate that the use of Delaunay variables provides a clearer insight in the essence of known canonical simplification algorithms, on the one hand, and makes their construction elementary, on the other. Finally, it is noteworthy that the transformation series required by the Delaunay variables approach provide a notable reduction in the number of terms to be evaluated when compared to the polar-nodal variables procedure. In particular, the number of terms of the generating function of the elimination of the parallax is reduced by about one third, while this reduction is of about two thirds for higher orders of the elimination of the perigee. Savings in the number of Hamiltonian terms are irrelevant in the elimination of the parallax, but reach the 60\% at the second of the elimination of the perigee, and an impressive 90\% at the fourth order.

\section*{Acknowledgemnts}

Part of this research has been supported by the Government of Spain (Projects AYA 2009-11896, AYA 2010-18796, and grant Fomenta 2010/16 of Gobierno de La Rioja).

\appendix

\section*{Appendix} \label{a:constants}

The integration ``constant'' $V_2$ is
\begin{eqnarray} 
V_2 &=&
\frac{n\,\alpha^4}{512a^2}\,\frac{J_2^2}{\eta^7}\left\{
\frac{\left(1-15 c^2\right)^2 \left(2-15 c^2\right)}{2(1-5c^2)^3}\,s^4\,e^4\sin4\omega
\right. \\ \nonumber && \left.
+\left[12\,\frac{6-43c^2+125c^4}{1-5c^2}
-(1-15c^2)\,\frac{25-126c^2+45c^4}{(1-5c^2)^2}\,e^2\right]s^2\,e^2\sin2\omega
\right\}.
\end{eqnarray}
The term $W_3$ of the generating function of the elimination of the perigee is
\begin{eqnarray}
W_3 &=&V_3-J_2^3\frac{\alpha ^6 n }{a^4 \eta ^{11}}\left\{ \frac{14-15s^2}{4-5 s^2}s^2  \left[
  \left(\frac{2925 s^8-7710 s^6+7064 s^4-2496 s^2+224}{2048(4-5s^2)^2}e^2
\right. \right. \right. \\ \nonumber && \left.
   -\frac{(14-15 s^2)\,(2-3 s^2)}{2048(4-5s^2)}s^2\right)e^2\sin2f+\left(-\frac{(14-15s^2)\,(2-3 s^2)}{512(4-5 s^2)}s^2
\right. \\ \nonumber && \left. \left.
   +\frac{6075 s^8-15960 s^6+14556s^4-5104 s^2+448}{1024 \left(4-5 s^2\right)^2}e^2
   \right)e\sin{f}
   \right]
\\ \nonumber &&
-\frac{(14-15 s^2)^2}{(4-5 s^2)^2}s^4\left[
\frac{135s^4-242 s^2+112}{2048(4-5 s^2)}e^3\sin (f+4 g)+\frac{3(2-3 s^2)}{2048}\,e^3 \sin (3 f+4 g)
\right. \\ \nonumber &&  \left.
+\left(\frac{30 s^4-55 s^2+26}{2048(4-5 s^2)}e^2+\frac{5(2-3s^2)}{2048}\right)e^2\sin(2f+4g)
+\frac{(2-3 s^2)}{4096}\,e^4 \sin (4f+4 g) 
\right]
\\ \nonumber &&
+\frac{3s^2 }{4-5 s^2}\left[
-\frac{14-15 s^2}{128}\,(2-3 s^2)^2\left(\frac{1}{8} e^2 \sin (4 f+2 g)+e \sin (3 f+2 g)\right)
\right. \\ \nonumber &&
-\left(\frac{3(14-15 s^2)}{1024(4-5s^2)}\,(85 s^6-22 s^4-96 s^2+48)\,e^2
\right. \\ \nonumber && \left.
+\frac{-3105 s^6+7251 s^4-5598 s^2+1424}{128}\right)e\sin(f+2 g)
\\ \nonumber &&
+\frac{(14-15 s^2)\,(-45 s^4-36 s^2+56)\,(2-3s^2)}{1024(4-5 s^2)}\left(\frac{1}{4} e^4 \sin (2 f-2 g)+e^3 \sin (f-2 g)\right)
\\ \nonumber && 
+\left(-\frac{3(14-15 s^2)}{4096(4-5 s^2)}\,(85 s^6-22s^4-96 s^2+48)\,e^4
\right. \\ \nonumber &&  \left. \left. \left.
+\frac{1890s^6-4317 s^4+3258 s^2-808}{512}e^2-\left(\frac{14}{16}-\frac{15}{16}s^2\right)\left(1-\frac{3}{2}s^2\right)^2\right)\sin(2f+2g)\right]
\right\}
\end{eqnarray}
where the integration constant $V_3$ is
\begin{eqnarray}
V_3 &=&-\frac{\alpha ^6 n J_2^3}{a^4 \eta^{11} \left(4-5 s^2\right)^3}\left\{
-\left(\frac{275}{16384}s^4-\frac{1445}{49152}s^2+\frac{317}{24576}\right)\frac{(14-15 s^2)^3}{(4-5 s^2)^2}e^6s^6\sin6\omega \qquad
\right. \\ \nonumber &&
   +\left[\left(\frac{225}{4096}s^6+\frac{2655}{32768}s^4-\frac{513}{2048}s^2+\frac{491}{4096}\right)\frac{(14-15 s^2)^2}{4-5 s^2}\,e^2
   -\frac{617625}{8192}s^8
\right. \\ \nonumber && \left.
   +\frac{4334325}{16384}s^6-\frac{2843175}{8192}s^4+\frac{826239}{4096}s^2-\frac{22429}{512}\right]e^4s^4\sin4\omega
   +\left[-\left( \frac{1164375}{8192}s^{12} \right. \right.
\\ \nonumber && 
   -\frac{8703375}{16384}s^{10}+\frac{6658575}{8192}s^8-\frac{2712045}{4096}s^6+\frac{644695}{2048}s^4-\frac{11333}{128}s^2
\\ \nonumber && \left. 
   +\frac{49}{4}\right)\frac{e^4(14-15s^2)}{(4-5s^2)^2}+\left(\frac{14338125}{2048}s^{12}-\frac{126428625}{4096}s^{10}+\frac{56930775}{1024}s^8
\right. \\ \nonumber && \left.
   -\frac{26745515}{512}s^6+\frac{6896989}{256}s^4-\frac{115555}{16}s^2+\frac{1575}{2}\right)\frac{e^2}{4-5s^2}+\frac{16407375}{4096}s^{10}
\\ \nonumber && \left. \left.
   -\frac{17059525}{1024}s^8+\frac{7037775}{256}s^6-\frac{1442021}{64}s^4+\frac{146825}{16}s^2-\frac{11883}{8}\right]e^2s^2\sin2\omega
\right\}
\end{eqnarray}

\begin{table*}[htbp]
\caption{Inclination polynomials $q_{i,j,k}$ in Eq. (\protect\ref{parallax}).}
\label{t:qijk}
\centering 
\begin{tabular}{ll}
\hline\noalign{\smallskip}
$q_{1,0,0}=1-\frac{3}{2}s^2$
\\ [1.5ex]
$q_{2,0,0}=\frac{5}{2}-\frac{21}{4}s^2+\frac{21}{8}s^4$
\\ [0.5ex]
$q_{2,0,1}=\frac{3}{4}-\frac{3}{4}s^2-\frac{15}{32}s^4$
\\ [0.5ex]
$q_{2,1,0}=-\frac{21}{8}+\frac{45}{16}s^2$
\\ [1.5ex]
$q_{3,0,0}=\frac{39}{2}-\frac{567}{8}s^2+\frac{2961}{32}s^4-\frac{315}{8}s^6$
\\ [0.5ex]
$q_{3,0,1}=\frac{87}{8}-\frac{837}{16}s^2+\frac{6813}{64}s^4-\frac{8145}{128}s^6$
\\ [0.5ex]
$q_{3,1,0}=-\frac{9}{8}-\frac{117}{16}s^2+\frac{2565}{256}s^4$
\\ [1.5ex]
$q_{4,0,0}=\frac{501}{2}-\frac{18909}{16}s^2+\frac{131157}{64}s^4-\frac{50049}{32}s^6+\frac{13815}{32}s^8$
\\ [0.5ex]
$q_{4,0,1}=\frac{3633}{16}-\frac{11961}{16}s^2-\frac{22509}{128}s^4+\frac{123309}{64}s^6-\frac{2596275}{2048}s^8$
\\ [0.5ex]
$q_{4,0,2}=\frac{783}{64}-\frac{13905}{128}s^2+\frac{26541}{64}s^4-\frac{277425}{512}s^6+\frac{1781595}{8192}s^8$
\\ [0.5ex]
$q_{4,1,0}=-\frac{40545}{32}+\frac{300525}{64}s^2-\frac{2956191}{512}s^4+\frac{2360115}{1024}s^6$
\\ [0.5ex]
$q_{4,1,1}=\frac{567}{16}-\frac{7533}{128}s^2-\frac{50409}{1024}s^4+\frac{136215}{2048}s^6$
\\ [0.5ex]
$q_{4,2,0}=-\frac{37611}{512}+\frac{10665}{64}s^2-\frac{384345}{4096}s^4$
\\
\noalign{\smallskip}\hline
\end{tabular}
\end{table*}

\begin{table*}[htbp]
\caption{Inclination polynomials $q_{i,j}$ in Eq.~(\protect\ref{perigee}). Terms $q_{i,0}=q_{i,0,0}$ are given in Table \ref{t:qijk}.}
\label{t:qij}
\begin{tabular}{@{}ll@{}}
\hline\noalign{\smallskip}
$q_{2,1}=(4-5s^2)\left(\frac{3}{4}-\frac{3}{4}s^2-\frac{15}{32}s^4\right)$
\\ [1.5ex]
$q_{3,1}=\frac{87}{2}-\frac{2109}{8}s^2+\frac{43551}{64}s^4-\frac{24705}{32}s^6+\frac{79425}{256}s^8$
\\ [1.5ex]
$q_{3,2}=s^2\,(14-15s^2)\left(\frac{63}{32}-\frac{2655}{256}s^2+\frac{8325}{512}s^4-\frac{2025}{256}s^6\right)$
\\ [1.5ex]
$q_{4,1}=\frac{3633}{4}-\frac{66009}{16}s^2+\frac{48645}{16}s^4+\frac{2187027}{256}s^6-\frac{7488675}{512}s^8+\frac{12896325}{2048}s^{10}$
\\ [0.75ex]
$q_{4,2}=\frac{783}{4}-\frac{19773}{8}s^2+\frac{882387}{64}s^4-\frac{584901}{16}s^6+\frac{50207085}{1024}s^8-\frac{33117525}{1024}s^{10}+\frac{68414625}{8192}s^{12}$
\\ [0.75ex]
$q_{4,3}=-s^2\,(14-15s^2)\left(\frac{441}{32}-\frac{10773}{128}s^2+\frac{76851}{512}s^4-\frac{25515}{512}s^6-\frac{91125}{1024}s^8+\frac{30375}{512}s^{10}\right)$
\\
\noalign{\smallskip}\hline
\end{tabular}
\end{table*}
\par


\begin{table}[htbp]
\caption{Inclination polynomials $p_{i,j}$ in Eq. (\protect\ref{normalized}).}
\label{t:qi}
\begin{tabular}{@{}ll@{}}
\hline\noalign{\smallskip}
$q_{1,0}=1-\frac{3}{2}s^2$
\\ [1.5ex]
$q_{2,0}=\frac{15}{32}(8-16s^2+7s^4)$
\\ [0.75ex]
$q_{2,1}=\frac{3}{8}(2-3s^2)^2$
\\ [0.75ex]
$q_{2,2}=-\frac{3}{4}+\frac{3}{4}s^2+\frac{15}{32}s^4$
\\ [1.5ex]
$q_{3,0}=45\left(14-\frac{1411}{16}s^2+\frac{29623}{128}s^4-\frac{9935}{32}s^6+\frac{107205}{512}s^8-\frac{7175}{128}s^{10}\right)$
\\ [0.75ex]
$q_{3,1}=\frac{135}{128}(2-3s^2)\,(4-5s^2)^2\,(8-16s^2+7s^4)$
\\ [0.75ex]
$q_{3,2}=-9\left(18-\frac{943}{8}s^2+\frac{21791}{64}s^4-\frac{32713}{64}s^6+\frac{98005}{256}s^8-\frac{28675}{256}s^{10}\right)$
\\ [0.75ex]
$q_{3,3}=-\frac{45}{32}(2-3s^2)\,(4-5s^2)^2\left(2-2s^2-\frac{5}{4}s^4\right)$
\\ [0.75ex]
$q_{3,4}=9s^2(14-15s^2)\left(\frac{7}{32}-\frac{295}{256}s^2+\frac{925}{512}s^4-\frac{225}{256}s^6\right)$
\\ [1.5ex] 
$q_{4,0}=-9\left(5005-\frac{655227}{16}s^2+\frac{8907105}{64}s^4-\frac{64836115}{256}s^6+\frac{134801885}{512}s^8 \right.$
\\ [0.5ex]
$\left.\hspace{1.8cm}-\frac{155889825}{1024}s^{10}+\frac{86809625}{2048}s^{12}-\frac{27768125}{8182}s^{14}\right)$
\\ [0.75ex] 
$q_{4,1}=-135(4-5s^2)$
\\ [0.5ex]
$\qquad\times\left(\frac{71}{2}-\frac{2137}{8}s^2+\frac{54865}{64}s^4-\frac{191509}{128}s^6+\frac{381165}{256}s^8-\frac{407365}{512}s^{10}+\frac{362775}{2048}s^{12}\right)$
\\ [0.75ex]
$q_{4,2}=45\left(366-\frac{44697}{16}s^2+\frac{535359}{64}s^4-\frac{3043877}{256}s^6+\frac{3362197}{512}s^8\right. $
\\ [0.5ex]
$\left.\hspace{1.8cm}+\frac{1185375}{512}s^{10}-\frac{1129325}{256}s^{12}+\frac{6075125}{4096}s^{14}\right)$
\\ [0.75ex]
$q_{4,3}=45(4-5s^2)$
\\ [0.5ex]
$\qquad\times\left(59-\frac{1761}{4}s^2+\frac{46233}{32}s^4-\frac{169331}{64}s^6+\frac{359527}{128}s^8-\frac{412985}{256}s^{10}+\frac{395775}{1024}s^{12}\right)$
\\ [0.75ex]
$q_{4,4}=45\left(27-\frac{1787}{16}s^2-\frac{21431}{64}s^4+\frac{692757}{256}s^6-\frac{3209999}{512}s^8 \right.$
\\ [0.5ex]
\hspace{1.8cm}$\left.+\frac{7209305}{1024}s^{10}-\frac{8072925}{2048}s^{12}+\frac{7246125}{8192}s^{14}\right)$
\\ [0.75ex]
$q_{4,5}=-63(4-5s^2)$
\\ [0.5ex]
$\qquad\times\left(\frac{3}{2}-\frac{5}{8}s^2-\frac{2467}{64}s^4+\frac{18115}{128}s^6-\frac{54075}{256}s^8+\frac{74775}{512}s^{10}-\frac{79125}{2048}s^{12}\right)$
\\ [0.75ex] 
$q_{4,6}=-9s^2\,(14-15s^2)\left(\frac{49}{32}-\frac{1197}{128}s^2+\frac{8539}{512}s^4-\frac{2835}{512}s^6-\frac{10125}{1024}s^8+\frac{3375}{512}s^{10}\right)$
\\
\noalign{\smallskip}\hline
\end{tabular}
\end{table}


\end{document}